# Large Language Model-Augmented Auto-Delineation of Treatment Target Volume in Radiation Therapy

Praveenbalaji Rajendran, Yong Yang, Thomas R. Niedermayr, Michael Gensheimer, Beth Beadle, Quynh-Thu Le, Lei Xing, and Xianjin Dai

*Abstract*— Radiation therapy (RT) is one of the most effective treatments for cancer, and its success relies on the accurate delineation of targets. However, target delineation is a comprehensive medical decision that currently relies purely on manual processes by human experts. Manual delineation is time-consuming, laborious, and subject to interobserver variations. Although the advancements in artificial intelligence (AI) techniques have significantly enhanced the auto-contouring of normal tissues, accurate delineation of RT target volumes remains a challenge. In this study, we propose a visual language model-based RT target volume auto-delineation network termed Radformer. The Radformer utilizes a hierarchical vision transformer as the backbone and incorporates large language models to extract text-rich features from clinical data. We introduce a visual language attention module (VLAM) for integrating visual and linguistic features for language-aware visual encoding (LAVE). The Radformer has been evaluated on a dataset comprising 2985 patients with head-and-neck cancer who underwent RT. Metrics, including the Dice similarity coefficient (DSC), intersection over union (IOU), and 95th percentile Hausdorff distance (HD95), were used to evaluate the performance of the model quantitatively. Our results demonstrate that the Radformer has superior segmentation performance compared to other state-of-the-art models, validating its potential for adoption in RT practice.

*Index Terms*—medical image segmentation, large language model, vision language model, radiation therapy.

## I. INTRODUCTION

RADIATION therapy (RT) is a widely used modality for the treatment of cancer [1]–[3]. In RT, ionizing radiation is used to kill the cancerous cells by inflicting damage to their DNA. The therapeutic efficacy of the RT is achieved by administering sufficient radiation doses to the tumor target while minimizing the exposure to normal tissues [4]. In RT, precision delineation of the treatment target plays an important role, and it directly impacts the treatment outcomes. Furthermore, advanced RT treatment plans, such as the volumetric-modulated arc therapy (VMAT), are more susceptible to contouring inaccuracies. However, manual contouring of target volume is a complex, laborious process, subject to intra- and inter-observer variations [5]. Moreover, studies have demonstrated that a fire amount of the manually delineated target volumes are subjected to changes during the peer review process [6]–[8]. Over the last decade, deep learning (DL) has achieved significant progress in medical image segmentation tasks. Convolutional neural network (CNN) has demonstrated significant achievements in medical image segmentation [9]–[16]. Among CNNs, UNet stands out as one of the most extensively employed networks for segmentation tasks. For instance, a UNet-based hybrid densely connected network has been proposed for hepatic tumor segmentation from the combination of magnetic resonance (MR) and computed tomography (CT) images[17]. Similarly, a three-branch two-dimensional (2D) U-Net termed multiple branch UNet (MB-UNet) has been proposed for the concurrent segmentation of the prostate and lesions from the T2-weighted, diffusion weighted (DWI), and apparent diffusion coefficient (ADC) MR images [18]. To facilitate rapid target contouring, U-Net has also been extended to 3D volumetric data. For instance, 3D-UNet has been proposed for segmenting the prostate and the related organs for dose optimization in radiation therapy [19]. Moreover, a two-channel input 3D-UNet has also been proposed for the target volume segmentation in head and neck cancer [20]. In another implementation, CUNet, a modified 3D U-Net with residual block integrated with attention center block, has been proposed for prostate segmentation from CT images [21]. Similarly, DSD-UNet, a 3D-UNet-inspired architecture incorporating residual connection and dilated convolution, has been implemented for target volume segmentation in brachytherapy [22]. However, these methods suffer from limited delineation accuracy and are deemed unacceptable in routine clinical practice.

Over recent years, attention-based transformers have significantly advanced natural language processing (NLP) and computer vision domains [23]–[28]. Moreover, transformer-based backbones have achieved comparable or better performance than those of CNN-based backbones. Inspired by their success, attention-based transformers have also been adapted for various medical segmentation tasks. For instance, CoTr model interleaves transformers between the CNN decoder and encoder to enhance the tumor and organ segmentation performance [29]. In another development, TransUNet employs stacked self-attention-based transformers as an encoder backbone for multi-organ segmentation from CT images [30]. Similarly, UNETR utilized a transformer-based encoder backbone for tumor and organ segmentation [31]. Recently, SWIN-UNETR, a more efficient U-shaped architecture incorporating hierarchical shifted windows-based transformers as the encoder, has been proposed for tumor and organ segmentation [32], [33].

The development of transformers also ushered the rise of vision language models (VLMs), where combined visual and text representation learning are utilized for various downstream tasks such as classification, object detection, and segmentation [34]–[36]. Notably, the segmentation of target objects described by the natural language expression has garnered much attention in computer vision, leading to the development of VLM architectures such as Vision-Language Transformer (VLT) [37], Referring image Segmentation using Transformer (ReSTR) [38], Language-Aware Vision Transformer (LAVT) [39], Clip-driven referring image segmentation (CRIS) [40], etc. However, the implementation of VLM for medical image segmentation is limited. In this study, we propose a VLM-based 3D medical image segmentation network called Radformer. The Radformer utilizes 3D medical images and text-rich clinical information to delineate the RT target volume. The Radformer employs a hierarchical shifted window (SWIN) transformer [33] as its backbone and integrates large language models (LLMs) comprising of Generative Pre-trained Transformer 4 (GPT-4) [41] and PubMed Bidirectional Encoder Representations from Transformers (PubMed-BERT) [42] to extract the text-rich clinical information. The language and visual information are integrated through language-aware visual encoding (LAVE) via the visual language attention module (VLAM) at each stage of the network. A CNN-based decoder is adopted to generate the 3D segmentation masks using the language-aware visual features. To evaluate the effectiveness of the proposed Radformer architecture, a public dataset comprising 2985 patients with head-and-neck cancer who underwent radiation therapy was used. Our Radformer achieves a Dice similarity coefficient (DSC) of 76%, an intersection over union (IOU) of 69%, and an 95th percentile Hausdorff distance (HD95) of 7.82 mm over a test set comprising 597 patients, respectively, significantly outperforms the state-of-the-art methods.

The summarization of this study is as follows:
1. Radformer, a VLM-based 3D segmentation approach utilizing the hierarchical vision transformer and LLMs has been developed for RT treatment target volume delineation.

2. The Radformer was evaluated on a large cohort of test data comprising 2985 patients to demonstrate its effectiveness and generalizability.

3. The Radformer outperformed the baseline 3D-UNETR by 15% with respect to mean DSC, and 17% with respect to mean IOU exhibiting high performance and excellent efficiency.

## II. MATERIALS AND METHODS

### A. Method

The overview of the Radformer is shown in Figure 1. The Radformer utilizes an encoder-decoder architecture, utilizing hierarchical vision transformers [33] in the encoder to generate cross-modal alignments between visual and textual information. A CNN-based decoder is utilized to generate 3D segmentation maps. In this section, we begin with the introduction of Language-Aware Visual Encoding (LAVE), followed by the Vision Language Attention Module (VLAM), the Language Gating Unit (LGU), and the CNN-based decoder.

#### 1) Language- Aware Visual Encoding (LAVE)

The encoder of the Radformer is designed to extract and fuse the features from the 3D images and the text-rich clinical data. Typically, a patient's clinical data comprises extensive unstructured information. To extract the features that are relevant to defining the treatment targets, we leverage the potential of the Generative Pre-trained Transformer 4 (GPT-4) [41]. We utilize tailored prompts to guide the GPT-4 to extract tumor-related information from the clinical data. A domain-specific Bidirectional Encoder Representations from Transformers (BERT) model pre-trained on the PubMed corpus called PubMed-BERT [42] was utilized to contextually embed the data extracted by GPT-4. The embedded high-dimensional word vector is represented as $L \in \mathbb{R}^{C_t \times T}$, where $C_t$ represents the number of channels (768, corresponding to the size of the PubMed BERT hidden layer) and $T$ represents the number of words.

Following the language feature extraction, we employ a SWIN-UNETR [33] based encoder to perform the combined visual feature fusion and visual language feature fusion. The encoder layer comprises four stages $i \in \{1, 2, 3, 4\}$. Each stage of the encoder has two transformer blocks, and it is designed to take two inputs, a 3D image of dimension $H \times W \times D$ ($V \in \mathbb{R}^{H \times W \times D}$) and vectorized tumor information of dimension $C_t \times T$ ($L \in \mathbb{R}^{C_t \times T}$). A patch partition layer is used to divide the 3D image volume into patches of dimension $\frac{H}{2} \times \frac{W}{2} \times \frac{D}{2}$ ($P \in \mathbb{R}^{\frac{H}{2} \times \frac{W}{2} \times \frac{D}{2}}$). The partitioned patches are then projected into an embedding space dimension $C_i$. In the encoder, the first three stages consist of two SWIN transformer blocks $\alpha_i$, a multimodal feature fusion module $\beta_i$, and a gating unit $\psi_i$. In these stages, the generation of semantic-aware visual features involves a three-step process inspired by language-aware vision transformer (LAVT) model [39]. In each of the initial three stages, the SWIN transformer blocks $\alpha_i$ utilize the features generated by the preceding stage as input to generate visual features as output $I_i \in \mathbb{R}^{C_i \times H_i \times W_i \times D_i}$. The resultant visual features $I_i$ are then integrated with the language features $L$ via a multimodal feature fusion module termed VLAM to generate multimodal attention features $M_i \in \mathbb{R}^{C_i \times H_i \times W_i \times D_i}$. A learnable gating unit $\psi_i$ called LGU weighs the attention features $M_i$; the weighted features $G_i$ are then combined elementwise to the visual feature $I_i$ to produce language-enriched visual features $E_i$. The final stage of the encoder comprises only the SWIN transformer block without the VLAM and LGU.

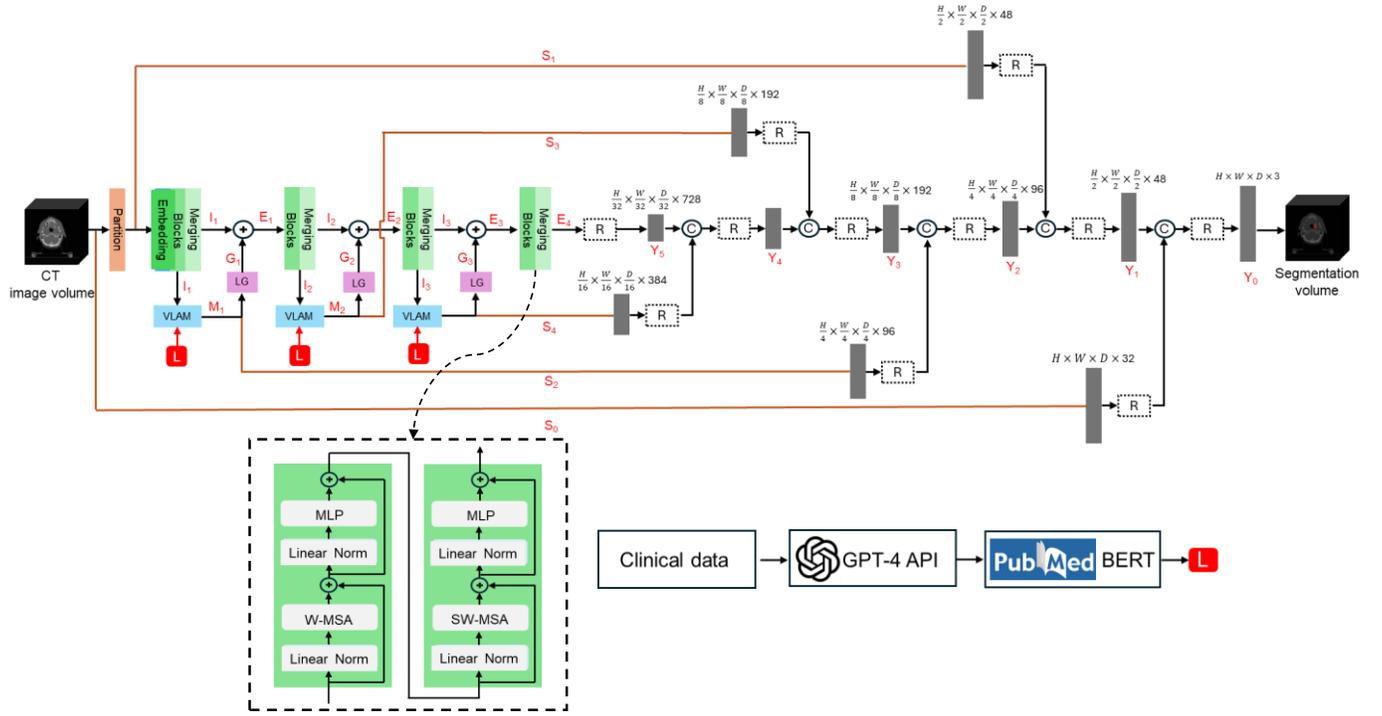

Fig. 1. Schematic of the proposed Radformer architecture. The Radformer utilizes 3D SWIN transformers as backbone to perform visual aware encoding. At each stage, the encoder generates visual features $I_i$ $i \in \{1,2,3,4\}$, which are used as the queries for the generation of position multimodal feature maps $M_i$ $i \in \{1,2,3,4\}$ through the visual langauge attention module (VLAM). These multimodal feature maps $M_i$ are then adaptively fused with the visual features $I_i$. The fused features $E_i$ are then used to generate the segmentation output through a CNN-based decoder employing skip connections.

### 2) Visual Language Attention Module (VLAM)

The schematic of the VLAM is shown in Figure 2. The VLAM utilizes the visual features as the query $I_i \in \mathbb{R}^{C_i \times H_i \times W_i \times D_i}$ and language features $L \in \mathbb{R}^{C_t \times T}$ as the key and value to generate position-specific sentence-level feature vectors $F_i \in \mathbb{R}^{C_i \times H_i \times W_i \times D_i}$ as follows.

$$I_{iq} = flatten(\rho_{iq}(I_i)) \quad (1)$$

$$L_{ik} = \rho_{ik}(L) \quad (2)$$

$$L_{iv} = \rho_{iv}(L) \quad (3)$$

$$F'_i = softmax\left(\frac{I_{iq}^T L_{ik}}{\sqrt{C_i}}\right) L_{iv}^T \quad (4)$$

$$F_{ij} = \rho_{iw}\left(unflatten(F'^T_i)\right) \quad (5)$$

Where, $\rho_{iq}, \rho_{iw}, \rho_{ik}, \rho_{iv}$, are projection functions. The projection functions $\rho_{iq}$ and $\rho_{iw}$ are implemented as $1 \times 1 \times 1$ convolutions followed by instance normalizations to generate projections resulting in channels $C_i$. The projection functions $\rho_{ik}$ and $\rho_{iv}$ are implemented as $1 \times 1 \times 1$ convolutions to generate $C_i$ number of output channels. The terms flatten and unflatten refer to the transformation of data from the original three spatial dimensions to a singular dimension and vice versa. The generated position-specific sentence-level feature vector $F_i$ is then fused with the visual projections $I_i$ to generate the multi modal feature maps $M_i$ as follows:

$$I_{im} = \rho_{im}(I_i) \quad (6)$$

$$M_i = \rho_{io}(I_{im} \odot F_i) \quad (7)$$

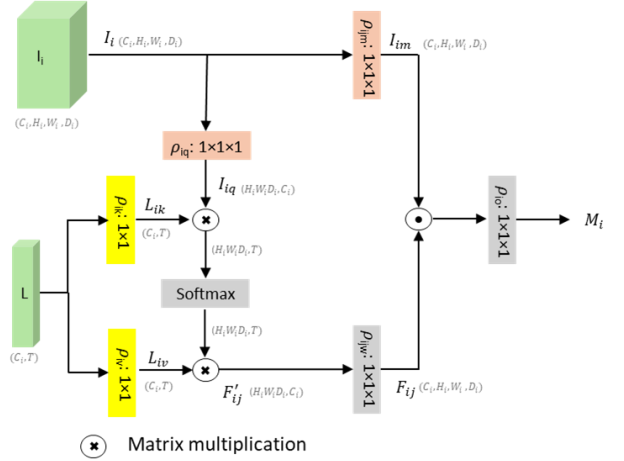

Fig. 2. Schematic of the Visual Language Attention Module (VLAM)

Where, $\rho_{im}$ and $\rho_{io}$ are projection functions and $\odot$

represents element-wise multiplication. These projection functions are implemented as $1 \times 1 \times 1$ convolutions followed by rectified linear unit (ReLU) activation.

### 3) Language Gating Unit (LGU)

Figure 3 illustrates the schematic of the LGU. The LGU is designed as a language gate that adaptively regulates the amount of information flowing to the subsequent stage of the transformer. The LGU is implemented as follows:

$$H_i = \gamma_i(M_i) \quad (8)$$

$$G_i = H_i \odot F_i + I_i \quad (9)$$

Where $\gamma_i$ refers to a two-layer perceptron; the first layer consists of $1 \times 1 \times 1$ convolution paired with RELU activation, and the second layer comprises $1 \times 1 \times 1$ convolution followed by hyperbolic tangent function. The $\odot$ in Eq. (9) signifies the element-wise multiplication.

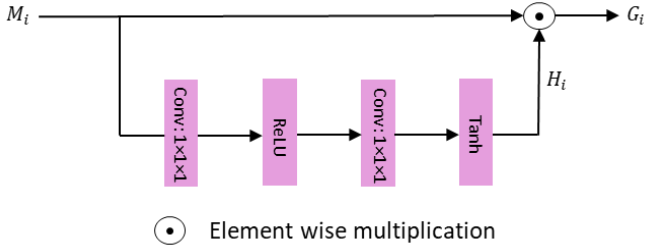

Fig. 3. Schematic of the Language Gating Unit (LGU)

### 4) Decoder

The multimodal feature maps $M_i$ generated by the encoder are utilized to generate the 3D segmentation maps using a CNN-based decoder. The decoding process at each stage of the decoder is implemented as follows:

$$S_0 = R(V), \quad V = V_1 + V_2 \quad (10)$$

$$S_1 = R(P), \quad P = P_1 + P_2 \quad (11)$$

$$S_j = R(M_j), \quad j \in \{2, 3, 4\} \quad (12)$$

$$Y_i = R[\upsilon(Y_{i+1}); S_i], \quad i \in \{0, 1, 2, 3, 4\} \quad (13)$$

$$Y_5 = S_5 \quad (14)$$

Where, R represents a residual block implemented with two $1 \times 1 \times 1$ convolutions with instance normalization. $\upsilon$ signifies the deconvolution based up sampling operation. The symbol [;] represents the concatenation among the channel dimensions. The final 3D segmentation outputs are generated from the feature maps $Y_0$ utilizing a $1 \times 1 \times 1$ convolution with a SoftMax activation.

### B. Data

We evaluated the Radformer on a public head-and-neck cancer dataset (RADCURE) [43]. The dataset comprises CT volumes and gross tumor volume (GTV) contours of patients who underwent radiation therapy, along with clinical data such as demographic details, clinical histories, and treatment specifics. The average age of the patients in the dataset was 63 years and the dataset consists of individuals diagnosed with oropharyngeal, larynx, nasopharynx, and hypopharynx cancer. The clinical information was based on the 7th edition TNM staging system and was standardized to adhere to the American Association of Physicists in Medicine (AAPM) Task Group report no.263 (TG263) nomenclature. To optimize the Radformer, we preprocessed the CT images and the GTV contours. The patient cases with data inconsistencies and corrupt labels were removed during the preprocessing. The CT images were center cropped to a size of $336 \times 336 \times 64$ voxels to ensure all regions of interest were included. The contours were used to generate segmentation masks, including the background and targets. The dataset was randomly divided into two groups: one comprising 2,388 datasets for training and validating the model, and another group with 597 datasets designated for testing.

The prompt developed for our experiments to guide GPT-4 in summarizing clinical information, along with two example summaries used for the patient cases, is shown in Figure 4.

| Prompt: |  |
|---|---|
| "Please infer and summarize the following patient's cancer information: ///{file_contents}/// in approximately 50 words, emphasizing the patient's age, gender, smoking history, cancer type, specific location, and overall stage. Exclude technical details like TNM classification or specific categories. For example, the resultant response should look like this ///A 62.6-year-old female ex-smoker has advanced oropharyngeal cancer located on the posterior wall. The tumor is large and invasive, affecting nearby lymph nodes significantly, but without spreading to distant parts of the body, categorizing it as stage IVB/// " | |
| Summarization: | |
| Example 1 | The patient is an 89.4-year-old male ex-smoker who has been diagnosed with early-stage laryngeal cancer, specifically located in the glottis. There is no evidence of the cancer spreading to the lymph nodes or other parts of the body, categorizing it as stage 0 |
| Example 2 | The patient is a 76.6-year-old male who is a current smoker. He has been diagnosed with stage III nasopharyngeal cancer. Despite the cancer affecting nearby lymph nodes, it has not spread to distant parts of the body |

Fig. 4. Prompt and examples of language Information used for patient cases.

### C. Training

#### 1) Loss Function

The loss function employed is a composite loss function comprising dice focal loss and Tversky loss from the MONAI library [44]. The composite loss function is formulated as:

$$L_{tot} = L_{DFC}(y_G, y_P) + L_{TV}(y_G, y_P) \quad (15)$$

Where, $y_G$ denotes the true label and $y_P$ denotes the predicted label. The dice focal loss, and Tversky loss are represented as $L_{DFC}$ and $L_{TV}$.

## 2) Implementation

Besides our introduced multimodal model, we also implemented advanced models including SWIN-UNETR and 3D-UNETR for comparisons. Each model (Radformer, SWIN-UNETR, and 3D-UNETR) was implemented using the PyTorch deep learning library [45] on an Nvidia RTX 4090 GPU with a batch size of 1. Adam optimizer with a learning rate of $4 \times 10^{-4}$ was utilized with a weight decay of $1 \times 10^{-3}$. All the models were trained for 100 epochs. We utilized HuggingFace's Transformer library [46] to integrate the PubMed BERT into the Radformer. The PubMed BERT utilized is a base model comprising of 12 layers with a hidden size of 768. The SWIN transformer layers of the Radformer are initialized with pre-trained weights from the BraTS dataset [32]. The remaining weights are randomly initialized.

### D. Evaluation Metrics

For evaluating the segmentation performance, we chose commonly used metrics including the dice similarity coefficient (DSC), intersection over union (IOU), and 95th percentile Hausdorff distance (HD95) [44]. These metrics were calculated on 3D volumes. Statistical analysis such as paired t-test was performed between our proposed method and other models across all metrics (DSC, IOU, and HD95).

## III. RESULTS

The performance of the Radformer was assessed utilizing the public RADCURE dataset as described in the above session. We also compared the performance of the Radformer with other state-of-the-art DL-based segmentation approaches including SWIN-UNETR and 3D-UNETR. The SWIN-UNETR and 3D-UNETR networks are DL approaches widely utilized for 3D segmentation tasks. As summarized in Table 1, the Radformer significantly outperforms other networks over all the metrics (p < 0.05). In particular, Radformer achieved a mean DSC score of 0.76, a mean IOU of 0.69, and a mean HD95 of 7.82 mm. The Radformer outperformed the baseline 3D-UNETR by 15% in terms of mean DSC and 17% in terms of mean IOU. Furthermore, a 45% improvement in the boundary accuracy (HD95) was noted compared to that of the 3D-UNETR.

TABLE I
COMPARISON OF THE PERFORMANCE OF RADFORMER FOR GTV SEGMENTATION WITH OTHER METHODS INCLUDING RADFORMER WITHOUT LAVE, SWIN-UNETR, AND 3D-UNETR

| Network | DSC | IOU | HD95 (mm) |
| --- | --- | --- | --- |
| Radformer (1) | **0.76±0.09** | **0.69±0.08** | **7.82±6.87** |
| SWIN-UNETR (2) | 0.69±0.11 | 0.64±0.09 | 12.88±6.60 |
| UNETR (3) | 0.66±0.09 | 0.59±0.07 | 14.28±6.85 |
| p-value (1 vs 2) | <0.05 | <0.05 | <0.05 |
| p-value (1 vs 3) | <0.05 | <0.05 | <0.05 |

The results of the ablation study containing the Radfromer implementation without LAVE are summarized in Table 2. Compared to the Radformer without LAVE (VLAM and LGU module), the Radformer with LAVE improves the segmentation performance by 8% in terms of mean DSC and 9% with respect to the mean IOU. Moreover, incorporating the language modules (VLAM and LGU) in the Radformer improved the segmentation accuracy (HD95) by 36%, with the improvements being statistically significant (p < 0.05).

TABLE II
COMPARISON OF THE PERFORMANCE OF RADFORMER FOR GTV SEGMENTATION WITH RADFORMER WITHOUT LAVE

| Network | DSC | IOU | HD95 (mm) |
| --- | --- | --- | --- |
| Radformer (1) | **0.76±0.09** | **0.69±0.08** | **7.82±6.87** |
| Radformer without LAVE (2) | 0.70±0.10 | 0.63±0.08 | 12.27±7.68 |
| p-value (1 vs 2) | <0.05 | <0.05 | <0.05 |

The qualitative results of the Radformer's GTV segmentation on a patient with early-stage laryngeal cancer and a patient with stage III nasopharyngeal cancer are illustrated in Figures 5 and 6, respectively. These figures are arranged into columns depicting the CT images, manual contours, and the GTV predictions by the proposed Radformer, Radformer without LAVE, SWIN-UNETR, and 3D-UNETR. Figures 5 and 6 show that the Radformer precisely segmented the GTVs in both cases when the language and image information were integrated. Whereas the Radformer without LAVE, the SWIN-UNETR, and 3D-UNETR overestimated the GTVs in most slices.

## IV. DISCUSSION

Delineation of the treatment target is a crucial step in RT treatment planning, and its accuracy significantly affects the quality of the RT plan. Currently, the delineation of target volume purely relies on human expert's manual process, which is time-consuming, laborious, and subject to intra- and inter-observer variability depending on the individual's knowledge and experience. Although DL-based segmentation approaches have made considerable advancements in recent years, its success in RT target delineation compared to normal organ segmentation is limited. At present, the delineation of RT treatment targets is still clinically unacceptable in most situations. The inferior performance of DL-based approaches for RT target delineation compared to normal organ segmentation could be attributed to the complexity of target delineation in RT. In fact, RT target delineation is a comprehensive medical decision-making process. In routine clinical practice, radiation oncologists consider the entire medical history and diagnosis of the disease when performing target delineation. In this study, we tested our hypothesis that, in addition to images, radiation oncologists leverage clinical data for RT target delineation. Therefore, integrating both text-rich clinical information and imaging data will improve DL-based auto-segmentation methods for RT target delineation.

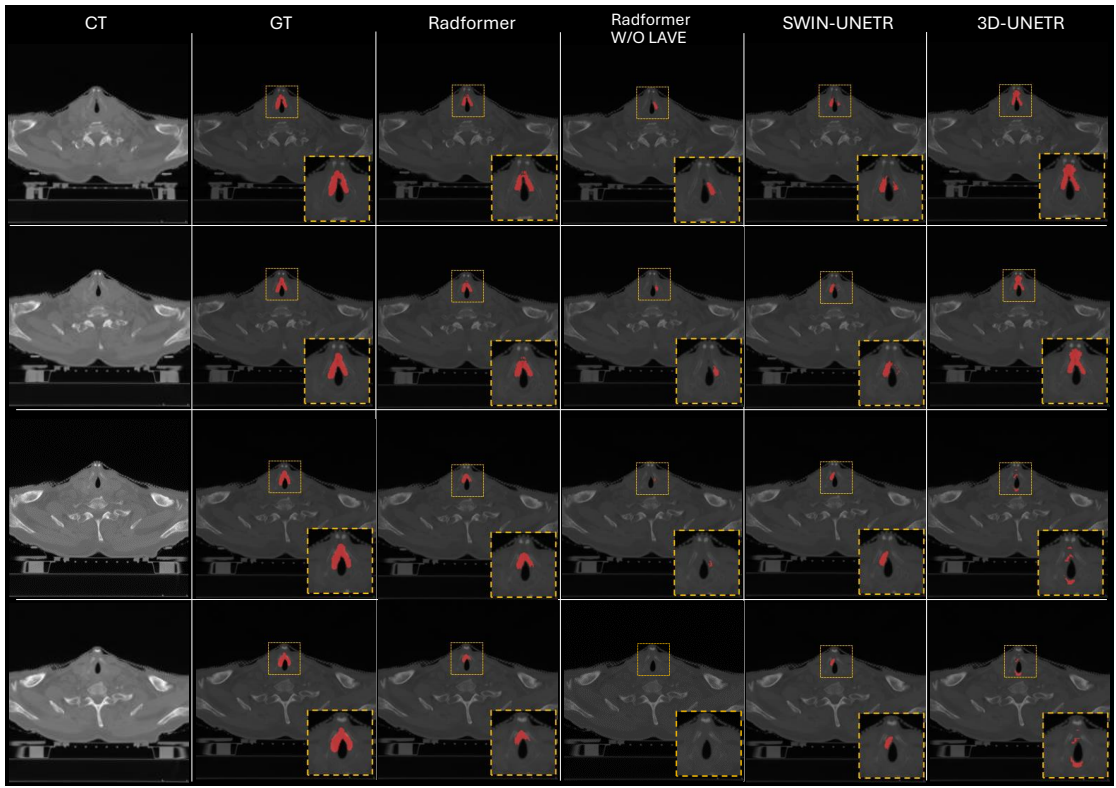

Fig. 5. Illustrative example of the GTV segmentation for a patient with early-stage laryngeal cancer

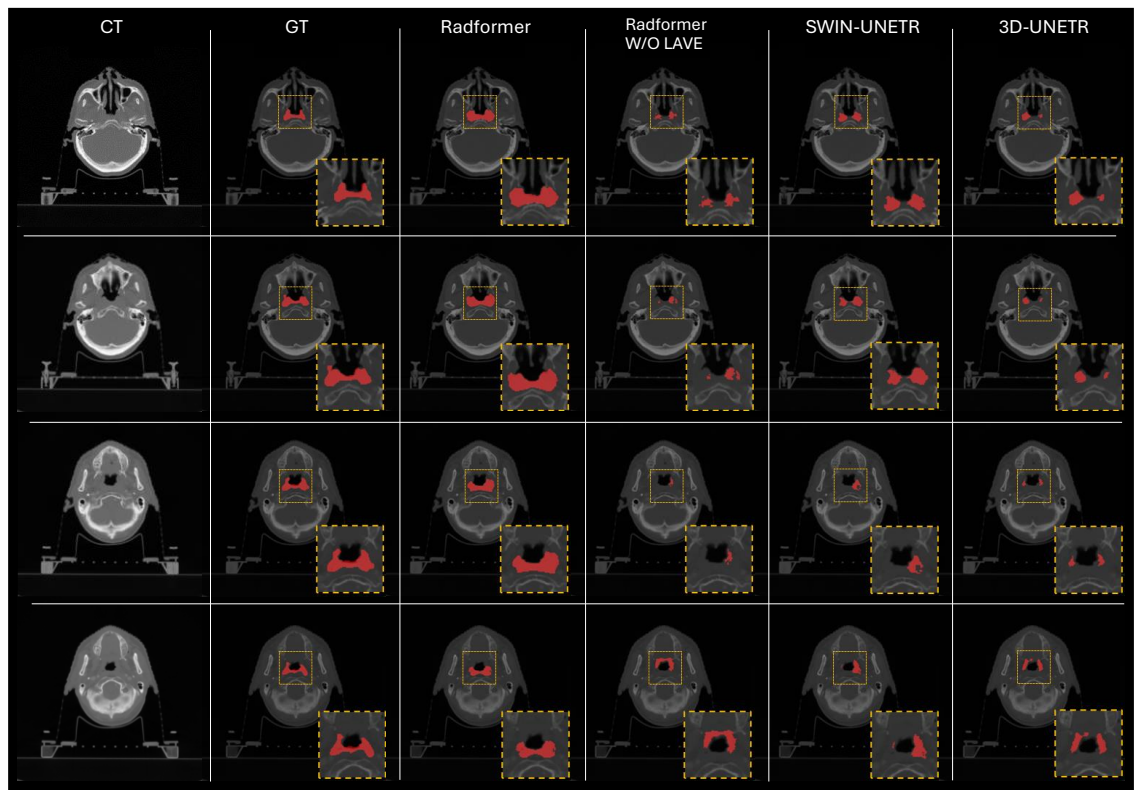

Fig. 6. Illustrative example of the GTV segmentation for a patient stage Ⅲ nasopharyngeal cancer

We developed a VLM-based 3D medical image segmentation network called Radformer. The Radformer utilizes the SWIN transformer as the backbone to extract visual features from images and LLMs to extract text-rich features from clinical data. The visual and linguistic features are fused through language-aware visual encoding and then leveraged by a CNN-based decoder to output segments. The Radformer was trained on a dataset from 2388 patients and evaluated on a test dataset comprising of 597 patients. Our results demonstrate that the Radformer achieves superior performance in comparison with the other state-of-the-art architectures purely leveraging visual features (SWIN-UNETR, and 3D-UNETR). Moreover, the statistical analyses using paired t-tests over the metrics DSC, IOU, and HD95 further substantiate the significant improvement achieved by Radformer. Furthermore, the enhanced accuracy in segmenting the GTV between the Radformer and Radformer without LAVE (Table 2) underscores the importance of linguistic and visual feature integration for GTV segmentation.

Figure 7 depicts the violin plots of segmentation performance distribution on all patient test cases, evaluated over the metrics including DSC, IOU, and HD95. These plots reveal the spread and density of the segmentation performance scores across ~600 patient cases. Figures 7(a) and 7(b) highlight that the Radformer architecture generally surpasses the performance of other models in most patient cases for both DSC and IOU metrics. This suggests a consistent, robust, and generalized segmentation capability. Furthermore, over other architectures such as Radformer without LAVE, SWIN-UNETR, and 3D-UNETR, it can be noted that the performance scores are tightly clustered within the interquartile range and are characterized by lower median values, indicating a relative underperformance in certain patient scenarios. This variability in performance substantiates the advantages of integrating linguistic and visual features. Furthermore, the violin plots representing performance over the HD95 metric indicate that the majority of the Radformer's performance is skewed towards a lower median value, suggesting enhanced accuracy in contour segmentation compared to other methods. This skewing towards lower values indicates the tendency of the Radformer to yield more precise segmentations with less deviation from the ground truth, as lower HD95 values correspond to smaller distances between predicted and actual contours. Moreover, the improvement in segmentation performance when linguistic information can be noted by comparing the performance distribution between the Radformer and Radformer without LAVE over all the metrics.

At present, the clinical data associated with the medical images are unstructured and extensive. In this study, we leverage the potential of LLMs to extract valuable information from the clinical data through customized prompts. In particular, we used GPT-4 to extract tumor-related information from clinical data. As GPT-4 only generates output texts for public access, we utilized PubMed BERT (an open-source model) to contextually embed the lesion information provided by GPT-4. Ideally, the process of extracting tumor-related information and contextual embedding can be masked by a single LLM. The proposed approach is generic and has the great

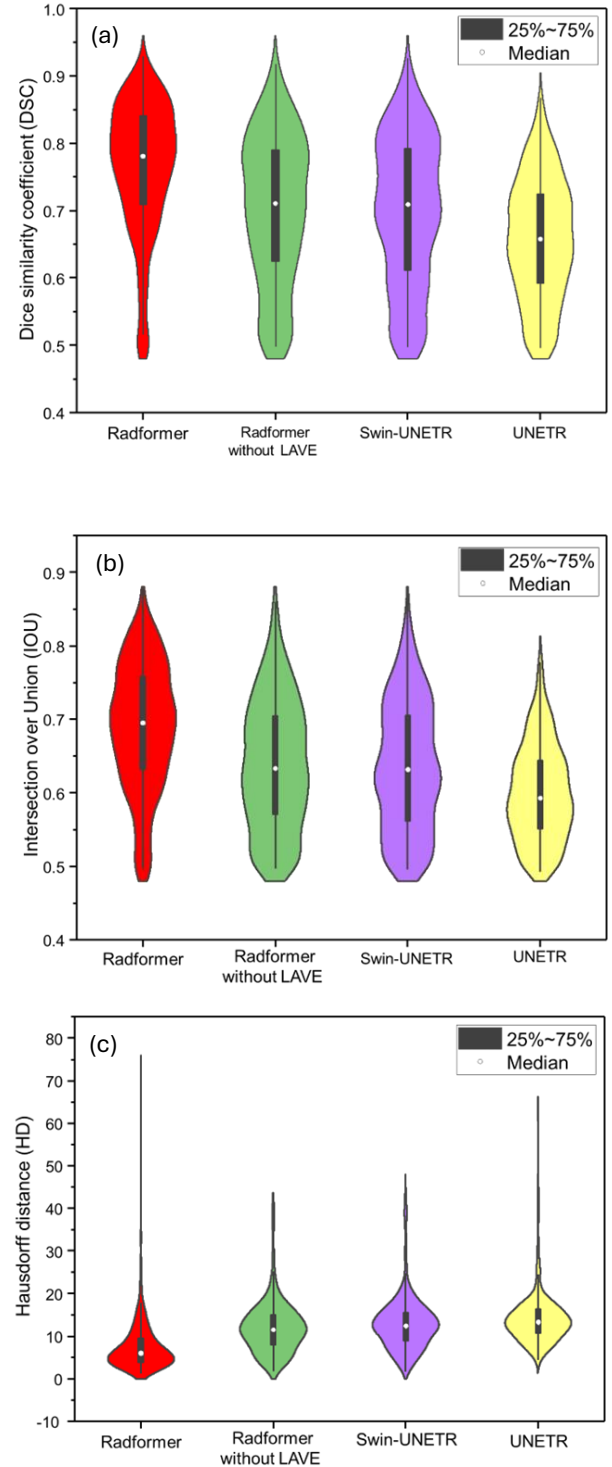

Fig. 7. Violin plots demonstrating the performance of the Radformer, Radformer without LAVE, SWIN-UNETR and 3D-UNETR on all patient test cases.

potential to be adaptable to the segmentation of various cancer types in radiation oncology.

Moving forward, integrating our proposed method into standard radiotherapy treatment planning could transform clinical practices. The Radformer's ability to delineate target volumes quickly and precisely could drastically decrease the manual effort and time typically required for this task. This increase in efficiency is especially important in the context of advanced RT treatment plans, where even small errors in contouring can lead to significant consequences.

## V. CONCLUSION

In conclusion, in this study, we propose a large language model-augmented approach for RT treatment target auto-delineation. Breaking away from the conventional architectures relying only on visual features, our model leverages text-rich clinical information and visual features to enhance the accuracy of target delineation. The proposed method could be adopted into routine radiotherapy treatment planning, offering a means to rapidly contour the target volumes with high precision.

## ACKNOWLEDGMENT

This research is supported in part by the Department of Defense Prostate Cancer Research Program Award W81XWH-19-1-0567, Stanford Radiation Oncology Prostate Cancer Research grant, Stanford Quality Incentive Metric Monies program, the Human-Centered Intelligence of Stanford University, and the National Institutes of Health (1R01CA223667, 1R01CA176553, and 1R01CA227713).